\documentclass[conference]{IEEEtran}
\IEEEoverridecommandlockouts
\usepackage{cite}
\usepackage{amsmath,amssymb,amsfonts}
\usepackage{algorithmic}
\usepackage{graphicx}
\usepackage{makecell}
\usepackage[breaklinks=true]{hyperref}
\usepackage{breakcites}
\usepackage{textcomp}
\usepackage{xcolor}
\usepackage[shortlabels]{enumitem}

\usepackage{footmisc}
\newlist{steps}{enumerate}{1}
\setlist[steps, 1]{leftmargin=1cm, label = Step \arabic*:}
\usepackage{listings}

\usepackage{amsthm}
\theoremstyle{definition}

\newtheorem*{remark}{Scenario}

\usepackage{booktabs}
\newcommand{\tabitem}{~~\llap{\textbullet}~~}

\usepackage[10pt]{moresize}

\def\BibTeX{{\rm B\kern-.05em{\sc i\kern-.025em b}\kern-.08em
    T\kern-.1667em\lower.7ex\hbox{E}\kern-.125emX}}
\begin{document}


\title{
Robonomics: The Study of Robot-Human Peer-to-Peer Financial Transactions and Agreements
{\footnotesize On Blockchains, Cryptocurrencies, Smart Contracts, and Decentralized Identity}
}

\author{\IEEEauthorblockN{Irvin Steve Cardenas}
\IEEEauthorblockA{\textit{Advanced Telerobotics Research Lab} \\
\textit{Department of Computer Science}\\
\textit{Kent State University}\\
Kent, Ohio, USA \\
icardena@kent.edu}
\and
\IEEEauthorblockN{Jong-Hoon Kim}
\IEEEauthorblockA{\textit{Advanced Telerobotics Research Lab} \\
\textit{Department of Computer Science}\\
\textit{Kent State University}\\
Kent, Ohio, USA \\
jkim72@kent.edu}


}

\maketitle

\begin{abstract}

The concept of a blockchain has given way to the development of cryptocurrencies, enabled smart contracts, and unlocked a plethora of other disruptive technologies. But, beyond its use case in cryptocurrencies, and in network coordination and automation, blockchain technology may have serious sociotechnical implications in the future co-existence of robots and humans. Motivated by the recent explosion of interest around blockchains, and our extensive work on open-source blockchain technology and its integration into robotics - this paper provides insights in ways in which blockchains and other decentralized technologies can impact our interactions with robot agents and the social integration of robots into human society.



\end{abstract}

\section{Introduction}
Beyond constrained task-driven human-robot collaboration, it is worthy to project future scenarios where a robot with even the utmost limited intelligence and autonomy might have the need to engage in relatively similar human social interactions, such as agreements and financial transactions.

Consideration on the concepts of agreements and financial transactions between humans and robots is not novel. Currently, human-robot interaction (HRI) research, for example, introduces such transactions in experiments that apply monetary rewards to game-like scenarios between robots and humans. But, such interactions fail to fully or accurately replicate the unmediated peer-to-peer financial transactions that take place between humans. For example, during in-lab experiments a human subject might expect that winning a monetary bet against a robot implies that in the end a human researcher will be the one making the monetary payment - not the robot itself. Although unexplored in current HRI research, this type of mediated interaction might present physical framing effects \cite{1997_colin_behavioral_game_theory} and even lead to expectation bias over such interaction. This could be the case in experiments that model competitive games between humans and robots and then apply traditional game theory to explain outcomes \cite{2016_sandoval_hri_bribery} \cite{Fraune_2019_is_hri_more_competitive_between_groups_than_between_individuals}. 

Hence, we further note that traditional game theory explores how rational agents play against other agents who are expected to be rational \cite{1944_von_neumann_theory_of_games_and_economic_behavior}. Humans are not over-rational agents and various features of an interaction, both physical and psychological, can impact the end-result of an interaction. Therefore, behavioral game theory  - which introduces psychology into game theory, might provide a better framework to assess such interactions \cite{1997_colin_behavioral_game_theory}. Nonetheless, this implies that we must further consider factors that might lead to predisposed biases, overconfidence, and other artifacts that lead to an unrealistic experience in a human-robot interaction. 

A human-robot interaction mediated by a third-party human, may be considered a factor that violates realism and which binds results to lab environments. After all, in a world where robots and humans co-exists we cannot, or perhaps should not, always expect a human to be present as a mediator. It is possible that the design of a human-robot interaction that more accurately replicates the direct interaction between a human and another human, e.g. during a bargain or a competitive game, might lead to different results that more accurately reflect human perspective on robots. Section \ref{perception-on-financial-transactions-contracts} highlights the difference between expectation and the reality of transactions related to bargains or of financial nature.

In HRI research that involves competitive games, bargains, and promises - we can partially attribute such unrealism to the constraints of conventional fiat money and the lack of tools to perform direct peer-to-peer exchanges of value between a robot and a human. Furthermore, we can observe that formal agreements between a human and a robot cannot be fully replicate nor incite the realism of a human-to-human interaction due to the lack of agency in robots, restrictions on the enforceability of such agreements, and overall the lack of technology that allows for these agreements to take place. For example, consider modeling the following agreement that incorporates two clauses that state that (a) if a robot with a given identifier \verb|alpha| completes a task \verb|beta| (e.g picking up Lego blocks from a given workspace) within time \verb|delta|, then (b) the robot \verb|alpha| will receive a reward in the amount of \verb|gamma| dollars or tokens. In this interaction, we might be interested in exploring the role and perception of a human as the enforcer or supervisor of a robot. But, similarly, we can reverse the roles and place the human as the agent that must perform the task, and the robot as the enforcer or supervisor. 

The \textbf{first scenario}, of a human as a supervisor, is novel and has not been explored. This is in part to previously stated constraints. But, we might still ask ourselves the following:

\begin{enumerate}[a)]
    \item How might one issue a  monetary reward to a robot? 
    \item Why might a robot need money?
    \item What utility does money bring to a robot?
    \item Where might a robot spend money?
    \item How can we make such agreement enforceable and provide stronger guarantees of fairness to the robot?
\end{enumerate}

The \textbf{second scenario} which places the robot as the supervisor is most commonly addressed in literature, particularly on behavior modification and persuasive robotics. The study of human psychology presents a system based on a token economy; tokens redeemable for goods or freedoms \cite{1967_behavior_modification_of_an_adjustment_class}. Persuasive robotics includes psychological rewards in addition to tangible rewards \cite{Siegel_2009_persuasive_robotics_the_influence_of_robot_gender_on_human_behavior}\cite{2019_assessing_the_effect_of_persuasive_robots}.
But, similarly, we might consider how tangible it is for a robot to offer a reward to a human, and how we can realistically model such interaction. We can explore the following questions, among others:

\begin{enumerate}[a)]
    \item How might a robot transfer a, monetary or non-monetary, reward to a human?
    \item How can a robot monitor a task assigned to a human?
    \item How can we make such agreement enforceable?
\end{enumerate}

As for the first scenario, we can consider a near future in which robots whether semi or fully autonomous might have the necessity to engage in financial transactions with a human. e.g.  a robot might find that it is most optimal to purchase a service as part of a global plan to fulfill a task - hailing a taxi instead of using its innate locomotion system. A robot might also be required to enter into a financial transaction as part of a proxied interaction for its respective owner (e.g. in telepresence). Hence, for robots, money could serve the purpose of a social tool as it serves humans, allowing for better engagement and interaction with a human society. 

But, the constraints of fiat money present a barrier to robots, not just the physical nature of money, but also the technosocial boundaries that are set by financial services over the ownership of money and the establishment of credit. For the most part, making an online purchase requires a credit card, and acquiring a credit card requires various human-centric processes such as an identification card, a credit score, or a physical address. The other question is over the enforceability of an agreement, how can we guarantee that a human will not cheat a robot? Or at least reduce the probability of such event. When considering either scenario with only human parties involved, we can address enforceability and fairness by relying on third-parties to mediate or arbitrate. 

For example, consider if a mother played the role of a supervisor and requested her child to perform a given task in exchange of a reward. If upon completion of the task, the mother denied the reward to the child on any given basis, the child can involve another party to serve as an arbitrator or mediator - we can imagine that another family member such as the father could play the role of arbitrator / mediator. But, we can also consider that for quantitatively measurable tasks we can introduce technology that monitors the task and provides trustworthy attestable information. Similarly, in agreements between a robot and a human, mechanisms should exists that not only introduce fairness, but also provide security over the obligations of not just the robot but as well of the human - concepts such as arbitrators and trusted sources of information should be available and applied.


Blockchain technology - smart contracts, cryptocurrencies, and the study of cryptoeconomics serve as the enabling constructs of the latter discussion. Beyond their application on agreements and peer-to-peer financial transactions between robots and humans, these technologies have deeper sociotechnical implications on the coexistence of robots and humans, the set of possible interactions between robots and humans, and also implications on the interactions between robots. Additionally, Decentralized Identity \cite{2019_decentralized_identity_foundation} and the ability of robots and things (IoT) to obtain such identity can be considered one of the interesting technologies derived from blockchain.

This paper describes future prospects of financial transacting and contracting between robots and humans, and amongst robots. The recipe for such interactions are based on three key technologies: (1) Blockchain(s), (2) Smart Contracts, and (3) Cryptocurrencies. The application of these technologies into the world of robotics leads us into uncharted territory that should be further explored by fields such as human-robot interaction, law, social policy and other social sciences. Section \ref{foundations} presents the foundations of this paper which include blockchains, cryptocurrencies, smart contracts and decentralized identity. Section \ref{integration-in-robotics} discusses the integration of such technologies into the world robotics. Section \ref{perception-on-financial-transactions-contracts} discusses a survey we performed and our current work. Section \ref{discussion} concludes the paper with a discussion that includes further considerations and additional applications.


\begin{table*}[htbp]\label{table:summary-of-properties-and-application}
\caption{Summary of Properties and Application in Robotics}
\begin{center}
\begin{tabular}{|c|c|l|}
\hline
\makecell{\scriptsize \textbf{Technology}} & \makecell{\scriptsize \textbf{Properties}} & \makecell{\scriptsize \textbf{Application in Robotics}} \\
\hline
\makecell{\scriptsize Blockchain}
& 
\makecell{\scriptsize Decentralized, Global, Immutable,\\ Verifiable, Secure \footnote{The previous properties make it secure}, Public \footref{footnote-public-blockchains}}
& 
\makecell[l]{
\tabitem \scriptsize Globally record references and authenticate access to robots and operational logs \\
\tabitem \scriptsize Blockchain addresses can serve as unique identifiers \\
\tabitem \scriptsize Use PKI to encrypt/decrypt information \\
\tabitem \scriptsize Serves as infrastructure for smart contracts \\ 
\tabitem \scriptsize Common infrastructure for decentralized identity \\ 
\tabitem \scriptsize Optional infrastructure for private/secure P2P messaging
} \\
\hline
\makecell{\scriptsize Cryptocurrencies \\ (Coins \& Tokens)} 
& 
\makecell{\scriptsize Decentralized, Global, Permissionless,\\Trusteless, Instantaneous, Irreversible, \\Secure, Programmable, Pseudonymous,\\ Controlled Supply }
& 
\makecell[l]{
\tabitem \scriptsize Can be used as currency or medium of exchange\\ 
\tabitem \scriptsize E.g. used to purchase (physical/digital) goods and services \\
\tabitem \scriptsize Tokens can represent fractional ownership of a robot (asset) \\ 
\tabitem \scriptsize Can be used to simulate token economy in behavioral studies \\ 
\tabitem \scriptsize Token can represent service rights to a robot (utility) \\ 
\tabitem \scriptsize Can be programmed through smart contracts
} \\
\hline
\makecell{\scriptsize Smart Contracts }
& 
\makecell{\scriptsize Decentralized, Global,\\Self-verifiable,\\ Self-executable,\\ Tamper-proof}
& 
\makecell[l]{
\tabitem \scriptsize Contract logic between robot and human can be defined \\ 
\tabitem \scriptsize Asset ownership can be represented as tokens and programmed\\ 
\tabitem \scriptsize Can model robot services and utilities as tokens \\ 
\tabitem \scriptsize Can monitor events on the blockchain and self-execute logic \\
\tabitem \scriptsize Access rights and consent to operation of robots can be defined
} \\
\hline
\makecell{\scriptsize Decentralized \\ Identity}
& 
\makecell{\scriptsize Decentralize, Global, Permissionless, Programmable \\
}
& 
\makecell[l]{
\tabitem \scriptsize Allows robots to attach historical records to an identity \\ 
\tabitem \scriptsize Allows robots to claim physical and digital property \\ 
} \\
\hline
\multicolumn{3}{l}{$^{\mathrm{1}}$The previous properties make it secure.}
\end{tabular}
\end{center}
\end{table*}

\section{Conceptual Foundation}\label{foundations}



\subsection{Blockchains}

Accounting/bookkeeping dates as far back as the times of Mesopotamia \cite{1963_keister_commercial_record_keeping_in_ancient_mesopotamia}, with the innovation of double-entry bookkeeping \cite{1985_aho_rhetoric_and_the_invention_of_double_entry_bookkeeping} serving to this day as the basic foundation of how we account for value. In modern times, this process has been digitized and the logic of debits and credits implemented on top of traditional database systems. At a high-level, the latter allows parties to have traceability of individual transaction and obligations, and can provide a global view of the entire state of accounts. Generalizing this concept, we can see that other industries also leverage processes and systems that allow them to trace transactions amongst various parties.

In the world of finance - banks and other third parties such as payment processors and clearing houses rely on the concept of maintaining a book of records (i.e. a ledger) of all financial transactions. In a fiat economy, an accurate version of ledgers is primordial to financial transacting. But, currently, the implementation of such ledgers is bespoke and centralized, with no transparency to external transacting parties - every party bound to a bespoke ledger. This in turn leads to inefficiencies related to data reconciliation and concerns when performing high-value network transactions.

A \verb|blockchain| offers itself as a solution to these problems by doing away with the need for centralized, bespoke, and opaque ledgers, and instead presenting the use of a single immutable, decentralized, publicly verifiable \footnote{\label{footnote-public-blockchains}In recent years the concept private and permissioned blockchains has emerged which by definition are not fully decentralized, and by design do not necessitate nor incorporate a native cryptocurrency. This papers argues for the use of a public blockchain.} ledger that can digitally \textbf{record anything of value} (e.g. a deed, a land title, identity, or a financial transaction). 

Historically, the genesis of the word and the concept of a blockchain was brought forth by Bitcoin, the fully-digital, decentralized cryptocurrency \cite{2008_nakamoto_bitcoin}. The blockchain is the underlying technology that enables Bitcoin and other decentralized platforms. Whereby by in Bitcoin, the blockchain is used mainly as a ledger that keeps track of all Bitcoin transactions \cite{2008_nakamoto_bitcoin}, in Ethereum its use is extended to include the storage, execution, and auditing of smart contracts \cite{2014_buterin_ethereum}. From a technical standpoint, we can consider that the actual data structure and the technologies underlying blockchains are not new. It is based on decades of academic research on topics that include cryptography, distributed systems, mathematics and incentive engineering. The true innovation lies in making all of these work together in harmony, and in the real-world deployment of a consensus protocol that combines these technologies - allowing users to cooperate on building together a chain of records all on their own, without the need of central party.

Since its inception, the technical architecture of a blockchain has evolved, and even novel architectures \cite{2016_jedusor_mimblewimble} are being implemented. But, the concept of a public ledger that allows anyone (or anything) to record and share valuable data in a secure, verifiable, tamper-proof way remains at the heart of all implementations\footref{footnote-public-blockchains}. It is what enables various use cases of cross-industry and global coordination and collaboration \footnote{Blockchain is being applied to various industries such as supply chain, finance, IoT, big data}. It is the technology that aims to eliminate the need for trust, and dependence on third-parties and intermediaries.








We leave it up to the user to further explore the technicalities of a blockchain by reviewing the ample literature such as the following \cite{2019_ieee_blockchain} \cite{2019_blockgeeks}, and instead we focus on discussing the implication of having a ledger with such properties available for use in the world of robotics (or in a world filled with robots). For the purposes of our work we focus on the use of public permissionless blockchains\footref{footnote-public-blockchains}, for we believe that the future is decentralized the concepts of trust and security should not lie in the hands of central authorities.


\subsection{Cryptocurrencies} \label{foundations-cryptocurrencies}

Cryptocurrencies provide a means to perform global financial transactions with anyone or anything that can generate a blockchain wallet address. Cryptocurrencies can be divided into two categories a native blockchain coin (e.g. Bitcoin, Ether, or Litecoin), or tokens. \verb|Coins| serve as a currency or medium of exchange, and transactions occur on their respective native blockchain. On the other hand, \verb|tokens| are a representation of a particular asset or utility that can be created through the use of smart contracts, e.g. commodities or reward points can represented as tokens. \textit{The interplay between a blockchain and smart contracts in essence makes cryptopcurrencies programmable money}.

To store coins or tokens a \verb|wallet address| is needed from the respective blockchain the agent will interact with. For the most part, the creation of an address such as an Ethereum address requires the derivation of a public key from a private key, and other cryptographic operations. For humans, user-friendly wallet software such as MetaMask \cite{2019_metamask} and Portis \cite{2019_portis} can simplify the creation of a wallet address and also simplify the interaction with decentralized applications (DApp) on blockchains like Ethereum. A \verb|DApp| is traditionally composed of a front-end application that connects to a smart contract. To send cryptocurrency or to interact with a smart contract, the address of the recipient or smart contract address is needed, then the sender's private key is used to sign the transaction; lastly the transaction is sent into the blockchain. The aforementioned wallet software solutions simplify these steps by displaying graphical user interfaces and performing the adequate cryptographic functions behind the scenes.

Various cryptocurrencies have previously existed, but Bitcoin is the first fully digital \textbf{decentralized} cryptocurrency to exist \cite{2008_nakamoto_bitcoin} - it's decentralized nature brought forth by the blockchain. Bitcoin, as a cryptocurrency, is generated through what is known as \verb|mining|. It is a process that secures the integrity of the blockchain by requiring a network of computers to validate transactions and compete to solve a cryptographic puzzle in order create a "block" of transactions and append it to the "chain", and in turn be awarded newly minted Bitcoins. These newly minted Bitcoins in essence serve as an incentive mechanism to keep the operators of these computers (\verb|miners|) honest and active. Other blockchains, such as Ethereum rely on the same process.

\subsection{Smart Contracts}
The idea of a smart contract was first conceived by Nick Szabo in the 1990's \cite{1994_szabo_smart_contracts_glossary}. In verbatim, Szabo proposed a "computerized transaction protocol that executes the terms of a contract" \cite{1994_szabo_smart_contracts_glossary}. In his work, Szabo essentially described a means by which trust on fallible parties can be transformed into functional trust provided by software that operates on predefined rules. Overall, smart contracts can be seen as digital agreements that can embody complex contractual relationships in code and which are self-executable, self-enforceable and tamper-proof.

Smart contracts were later popularized by Bitcoin and fully put into practice by Ethereum \cite{2008_nakamoto_bitcoin} \cite{2014_buterin_ethereum}. In bitcoin, simple implementations of smart contracts were generated using a stack-based scripting language that allowed users to program Bitcoin transactions. Such programs allowed for execution of multi-signature transactions or escrow payments. In later years, Ethereum was developed as an effort to create a world computer, that leverages a blockchain and which provides a platform for developing and deploying smart contracts. Unlike Bitcoin, Ethereum provided a Turing-complete programming language that allows for more complex smart contracts.

The fully digital, and decentralized, nature of smart contracts and cryptocurrencies allows humans and robots to develop and enter into unilateral and bilateral agreements, as well as engage into peer-to-peer (unmediated) transactions. 

\subsection{Decentralized Identity}
Very few concepts are more important to functioning in a human society and economy than the concept of identity. As humans, we make use of our identities to establish trust and make claims of ownership or individual rights. Presently, human identity can be claimed through passports, driving licenses, or by context-specific identifiers such as a work ID or even a social media account. Tied to these identities are the attributes that define and make a human identifiable. Attributes such as a birth date, a name, an address, biometric information, and unique identifiers issued by government entities as passport numbers or driver's license numbers. These attributes are part of our identity. Often than not, a history of financial transactions and obligations are tied to identities and used to qualify an individual (e.g. credit worthiness). In future social road maps, in countries like China, a social credit system is even tied to government-issued identifiers \cite{2019_chinese_social_credit}.


\verb|Decentralized Identity| centered around the standard of self-sovereign identifiers reverses the traditional paradigm of identity \cite{2019_w3c_decentralized_identifiers}. Fundamental to the Decentralized Identity paradigm is allowing users to directly create, manage and dispose of many decentralized identifiers (DIDs) without the need of third parties, while still being able to recall the properties attached to such identifiers. \cite{2019_w3c_decentralized_identifiers} \cite{2019_decentralized_identity_foundation}. This approach of giving the user full control is what makes it \textit{self-sovereign}. To this end, the use of a decentralized infrastructure and methods, such as a blockchain and public-key cryptography, are suitable. For example, instead of identity distribution authorities, DIDs could be issued and access control could be handled by a blockchain. But, it is worthy to note that a public key derived from a private key can suffice as a unique identifier, but we would require a proper means to record and trace transactions tied to such identifier - the blockchain provides the necessary infrastructure for such transactions.

This paradigm has as many implication on humans, as well as on robots. For humans, this means the possibility to have more control over one's identity. For robots this means the possibility to even have an adequate means of identification and the ability to attach historical information/attributes that represent an identity. 



\section{Robot-Human Transactions}\label{integration-in-robotics}
We consider the following: (1) A blockchain can serve as a ledger on which robots and humans may access and record anything of value, such as an ownership title or a financial transaction. (2) Smart contracts can encode agreement logic between a robot and a human that is self-enforceable self-verifiable. (3) Cryptocurrencies can allow robots to hold financial obligations and enter into exchanges of value with a human, and vice versa. Lastly, we consider that the application and extension of the three can lead to more sophisticated and realistic interactions between robots and humans; and in the near future their application could ease the integration of robots into society. Table \ref{table:summary-of-properties-and-application} presents a summary of the properties and application of each technology in robotics. In the following sections we elaborate their applications.

\subsection{A decentralized, immmutable and publicly verifiable ledger}

Similar to its application in IoT\cite{2017_kshetri_blockchain_strenghten_overview}, a blockchain can be used for security and access management schemes of robots, data, and communication. Public key encryption can be used to encrypt messages between robots and data logs that are stored externally. Digital signatures can be further used to authenticate access to the data and the access to control a robot. If low-latency and low-cost transactions are possible, a blockchain may also be used for "\verb|on-chain|" communication and signaling. Agents may execute transactions on a given blockchain address and other agents may listen to transactions on such address. For such scheme, a blockchain that provides sub-second confirmation time such as \cite{2018_thundercore_protocol}, may be most suitable. Other proposed communication protocols such as Whisper \cite{2014_buterin_ethereum} can also be leveraged.


\subsection{Contracts} \label{contracts}





Smart contracts can encode the logic of unilateral\footnote{One-sided agreements in which only the offeror promises to fulfill a given obligation after the occurrence of a given act \cite{2019_cornell_unilateral_contract}.} and bilateral\footnote{Two-sided agreements in which a promise is given in exchange for a promise in return. Each party is both an obligor - a party bounded to another party, and an obligee - a party to whom another is obligated or bound \cite{2019_investopedia_bilateral_contract}.} legal agreements. Their digital nature allows cyberphysical agents to interact with such logic and enter into agreements with other robots or human agents.

\begin{figure}[htbp]\label{fig:autonomous-vehicle-dao}
\centerline{\includegraphics[width=0.4\textwidth]{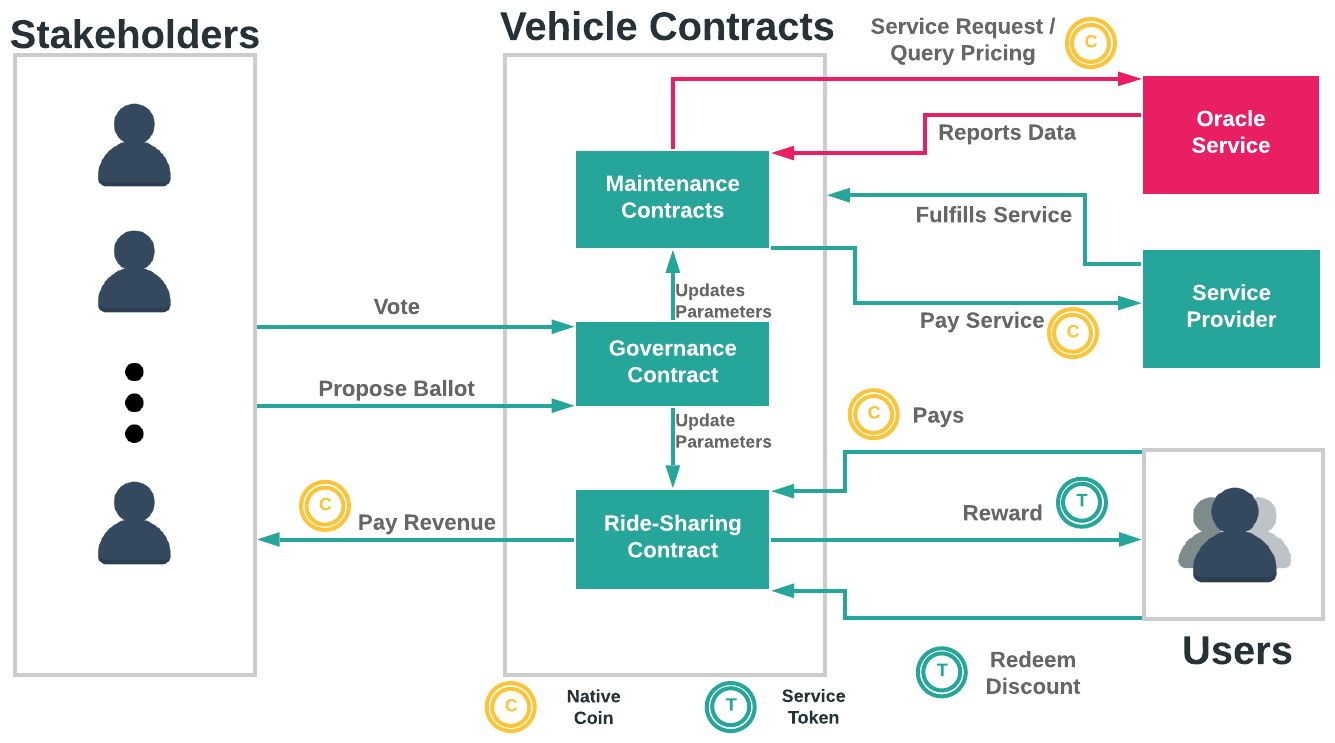}}
\caption{Autonomous Vehicle Smart Contract Example}
\end{figure}

\subsubsection{Ownership}
Listing \ref{lst:vehicle_ownership_contract} shows the partial interface definition of a smart contract for the use case in Fig. \ref{fig:autonomous-vehicle-dao}. It presents the use case of fractional ownership of an autonomous vehicle. The vehicle in turn is used as an autonomous taxi. Rather than tokenizing the ownership, the smart contract simply defines a set of owners in line 4, and a function that allows ownership transfer in line 10. A more extensive implementation of such smart contract, would include other smart contracts that automate additional processes and allow for on-chain governance of the ride-sharing service.

A governance contract could allow owners of the vehicle to propose ballots to update parameters of other smart contracts. A ballot could be proposed to increase the service fee for taking a ride on the vehicle. Another ballot could be proposed to increase the allocated funds for maintenance of the vehicle. In Fig. \ref{fig:autonomous-vehicle-dao}, such maintenance contract could be triggered by the vehicle when sensors indicate it needs servicing (e.g. an oil change). The contract can communicate to an \verb|Oracle| service that can provide pricing information of nearby service providers. A service provider could then be informed by the Oracle service that the vehicle is in need of service and generate an escrow contract that includes a pricing quote and a set of obligations that the provider will meet.

The vehicle can then accept the contract and send the amount quoted in cryptocurrency to the escrow contract. Upon completion of the service, the service provider signals the escrow contract that the service has been completed. Then the vehicle is notified and can assess whether the contract/service agreement has been fulfill (e.g. checking oil sensors). If the criteria is met, the vehicle can confirm the service by signing a transaction that is sent to the escrow contract. The escrow contract would then release the funds to the service provider.



\lstset
{ 
    basicstyle=\scriptsize,
    numbers=left,
    stepnumber=1,
    showstringspaces=false,
    tabsize=1,
    breaklines=true,
    breakatwhitespace=false,
    xleftmargin=0.7cm
}
\begin{lstlisting}[caption={Partial Interface Definition of Smart Contract}, captionpos=b,
label={lst:vehicle_ownership_contract}]
contract RideSharingContract {
 address public vehicle;
 bytes32 public vin;
 address[] public owners;
 address passenger;
 
 function vin()constant returns(bytes32 vin);
 function owners()constant return(address owners);
 function setOwners(address [] _ownerAddrs);
 function approveTransfer(address _to);
 function requestRide();
 
 event RideReq(address indexed _passengerAddr, uint256 rideCost);
 event Transf(address indexed _from, address indexed _to);
 event Appr(address indexed _owner, address indexed _approved);
}
\end{lstlisting}

\begin{figure}[htbp]\label{fig:unilateral-contract-flow}
\centerline{\includegraphics[width=0.4\textwidth]{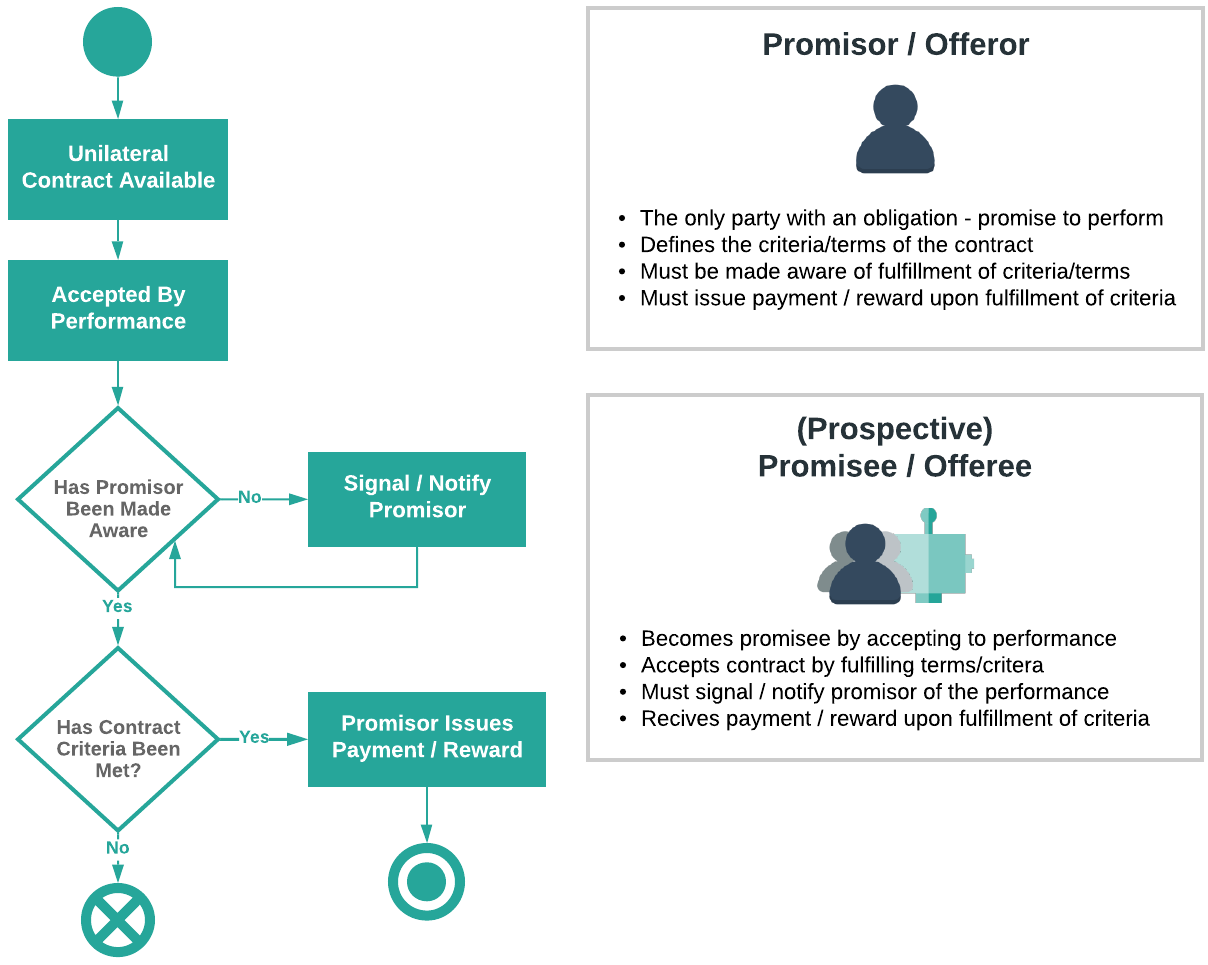}}
\caption{Unilateral Contract Flow}
\end{figure}


\subsubsection{Escrow Services}
Figure \ref{fig:unilateral-contract-flow} presents a state diagram of a unilateral contract. One of the most common examples are open requests or reward contracts, e.g. offering a reward for a lost pet. By offering a reward, the owner promises to fulfill the payment should anyone fulfill the obligation of returning the pet. Similarly, a robot can offer an unilateral contract to the public. In \cite{2018_cardenas_hri_robots_blockchain}, the robot can either offer a unilateral or bilateral contract to any individual willing to transport the robot to a given destination. In the unilateral version of the contract, the robot initializes the contract by defining its current location based on its GPS location, the goal destination and the payment for taking the robot to the goal destination. A human willing to accept the contract can carry the robot to the destination and sign the smart contract upon completion.


Legal contracts are backed by a legal system and are subject to interpretation. In smart contracts "the code is law". That is, the logic defined in a smart contract and the inputs provided to the smart contract define the ultimate outcome. Hence, the logic of a smart contract must be carefully drafted and the inputs provided to a smart contract must be trustworthy. 


\begin{figure}[htbp]\label{fig-escrow-agents}
\centerline{\includegraphics[width=0.5\textwidth]{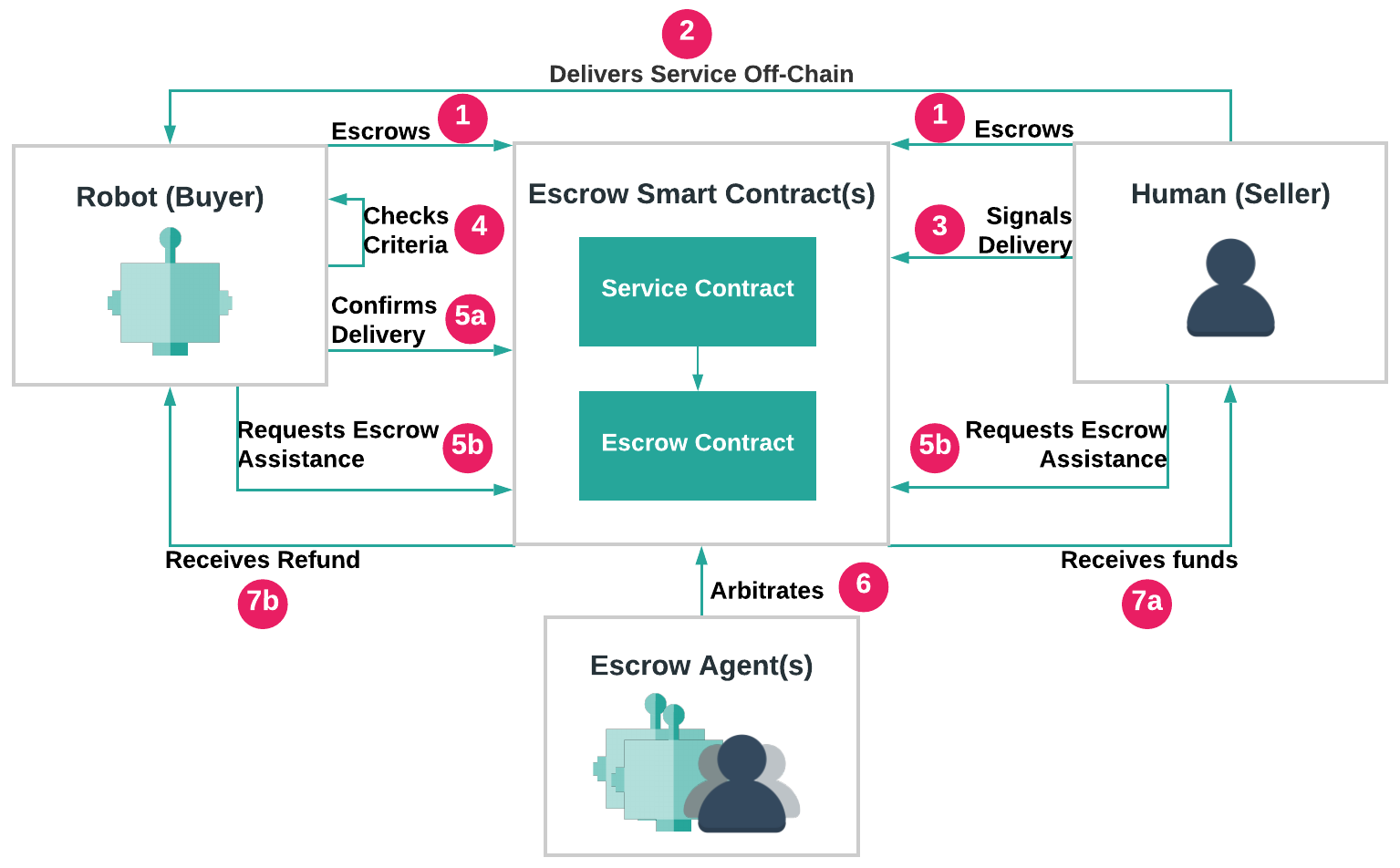}}
\caption{Decentralized Escrow Service Agents}
\label{fig}
\end{figure}

Meaning that conditional logic, such as the transfer of funds, will be executed given the respective input. Hence, we must consider what a breach of contract would entail, and how a robot could file for a breach of contract on a human, and vice versa. We can consider that a \textit{"breach of smart contract"}, in the human-sense, could be brought upon either by (a) a malicious actor that gains access to the private keys of either party, (b) a malfunction of the robot, or (c) an error from a data source or a malicious data source provider. Code bugs can also cause issues, but can be prevented with proper auditing.


To deal with such issues, in a buyer-seller scenario a smart contract such as in \cite{2019_solidity_safe_remote_purchase} can be used. But, we can also consider the use of an escrow smart contract that allows an escrow agent to arbitrate and that uses an oracle service to provide trustworthy data to the smart contract. In a decentralized escrow system, the parties transact through an escrow smart contract in which an escrow agent chosen by both parties oversees the transaction. If a dispute over the transaction occurs, the escrow agent can intervene and choose whether either party deserves to be refunded or paid. Figure \ref{fig-escrow-agents} illustrates the flow of an escrow transaction. 

\subsection{Human-Robot Interaction}

\begin{figure}[htbp]\label{fig:sequence-diagram-chess}
\centerline{\includegraphics[width=0.4\textwidth]{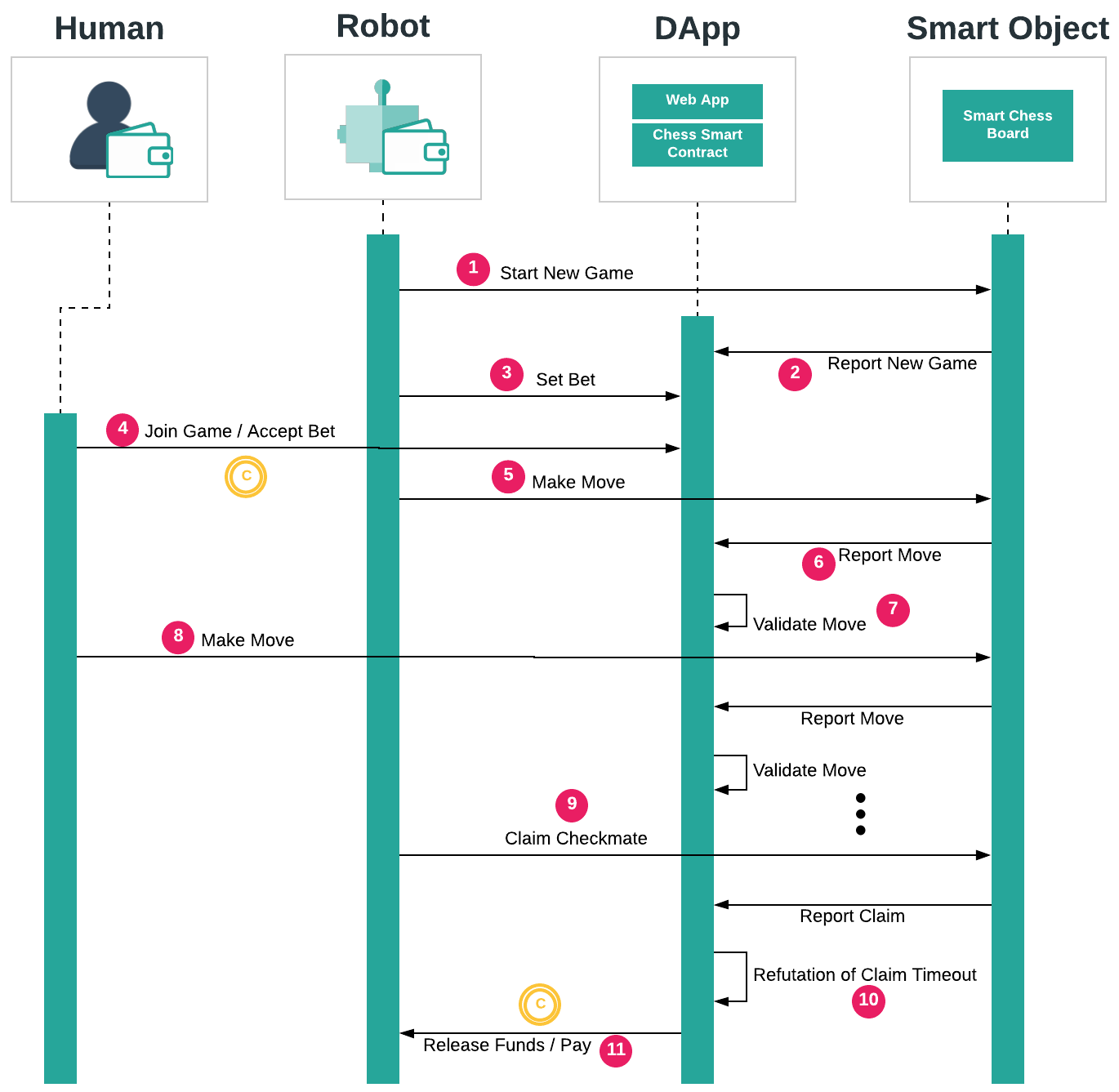}}
\caption{Sequence Diagram: Human vs Robot Chess Game + Betting}
\end{figure}

\subsubsection{Games} Given the peer-to-peer nature of the technology we can more faithfully design ummediated interactions that allows a robot to directly engage into a bilateral agreement with a human. Figure \ref{fig:sequence-diagram-chess} shows the sequence diagram of a physical game of chess between a robot and a human. Both parties make a traditional bet of winner takes all.

A traditional implementation of this human-robot interaction within a lab environment would require that a human researcher initially present the human subject with the context of the game. Essentially, presenting a competitive game with a monetary reward at stake. The winner of the game receives the monetary reward. The monetary compensation is often defined as a rewards card. Considering framing effects as discussed in \cite{1997_colin_behavioral_game_theory}, this game does not fully lend itself to a realistic interaction that we might expect to take place between a human and a robot with a high sense of intelligence and agency. We might expect that during a competitive game, a human subject might not only consider what's at stake but also consider how the opposing party views what is at stake (e.g. utility of the award).

In the context of a rewards card, we cannot realistically expect that a robot has any personal utility for the card due to the fact that a robot cannot personally access traditional physical, or online marketplaces, and place an order with such rewards card. As noted earlier, our society's traditional financial infrastructure is inadequate for cyberphysical agents. On the other hand, as \cite{2018_cardenas_hri_robots_blockchain} discusses, it is more feasible for a robot agent to use cryptocurrency to financially transact with humans and other robots. Given the emergence of decentralized marketplaces - marketplaces that do not leverage traditional financial infrastructure or have stringent know-your-customer (KYC) requirements, and often deal solely with cryptocurrencies - we can imagine that after winning a bet a robot can purchase a digital good such as a non-fungible token \cite{2018_blockgeeks_erc721_cryptokitty}, a piece of virtual property on a blockchain game, or purchase a physical good or service through other decentralized marketplaces such as OpenBazaar \cite{2019_openbazaar}.

Figure \ref{fig:sequence-diagram-chess} highlights the important steps in the sequence as dark pink circles embedded with numbers. Four actors are presented, the human agent, the robot agent, a DApp and a smart chess board. Both the human agent and the robot agent have a wallet address and/or wallet software. They make moves by physically moving the chess pieces on a board. The chess board is deemed smart because it uses a monitoring system and a set of physical buttons to report back to the smart contract that models the chess game. In future scenarios, we can imagine that a chess board and other physical electronic games would natively connect to a blockchain. 

The smart contract validates the moves and is in charge of keeping all bets, only releasing funds to the wallet address of the winning player. The DApp is a user interface connected to the chess game smart contract. It provides visual feedback of the state of the game and also allows either player to sign the contract and place a bet. A human user can simply scan a QR code on the application which dictates the rules of the game and amount of cryptocurrency required to join the game. A robot can use the latter, or directly send a transaction to the chess smart contract using a library such as Web3 in Ethereum.


Other games could be similarly modeled, that further engages the robot as the verifying party of the transaction in the game. That is, requiring the robot to approve whether a move is valid, or whether the game has been won or lost by the opponent. Hence, the concepts of cheating and greed can be more realistically modeled since the robot would be the actual agent considered in charge of performing or approving a transactions (not a third-party human). Another gaming interaction is presented in \cite{2019_cardenas_blockchain_game}. The latter presents a racing / battle game in which remote human drivers get to stake cryptocurrency or purchase the right to drive a robot and race against each other, additionally allowing the entire world to make bets against either robot and to purchase additional "power ups" for the robots. This is a novel cyberphysical blockchain game that leverages economic incentives and cryptographic functions to allow for globally cooperative and adversary gaming. 

\subsubsection{Commitment Strategies and Conditioning}
We can further consider the use of smart contracts to study commitment strategies and conditioning \cite{1955_strotz_myopia_commitment_strategy} \cite{1972_mischel_self_control_marshmallow_oreo} \cite{1975_ainslie_self_control_pigeons}. We can imagine the interplay of smart contracts and sophisticated machines that can counteract a human's short-term impulses by entering into agreements with a human. For example, a smart fridge can enter into a temporal agreement with a human and prevent the human from opening the fridge late at night or access to certain food cabins. Placing such logic on a smart contract, once again, means the agreement is self-enforcing, self-verifiable, tamper-proof - unstoppable.
From here on, the list of possible interaction is plenty. A thorough list and implementation of such interactions are presented in \cite{2019_cardenas_ro_man_workshop} as part of a workshop presented at the 2019 RO-MAN conference.

\section{Perception on Financial Transactions and Contracts with Robots}\label{perception-on-financial-transactions-contracts}

Drawing upon early work on behavioral economics, we presented a series of survey questions to infer on beliefs regarding robot-human financial transactions and agreements. Each survey scenario/questions captured, on average, responses from 50 college participants ages 18 - 28. We presented the same surveys to 20 participants with knowledge in cryptocurrencies and blockchain technology. Full details on the survey are presented in \cite{2019_cardenas_ro_man_workshop}. In this section we highlight one of the open-ended scenarios, summarized as follow:

\begin{remark}
Suppose you are living in the city, in a not so distant future where robots are common.  On a lazy Saturday you decide to make a trip to the grocery store. As you walk towards your vehicle to drive to the grocery store, a service robot approaches you. The service robot kindly asks you for a ride to a location near the grocery store. It explains that its power supply is malfunctioning and needs to arrive to the location as soon as possible to complete a task. In exchange, the robot offers to pay you a reasonable amount of money for your troubles.
\end{remark}

Participants were asked a set of questions that included whether they would accept the robot's offer, and the method of payment that they envisioned the robot using. In this scenario, 84 percent of the participants considered it "extremely likely" to accept the robot's offer \footnote{A 5-point scale was used that ranged from "Extremely Likely" to "Not at all Likely"}. In terms of payment method \footnote{Five options were given: (1) Cash / Fiat Money, (2) Credit Card, (3) Check, (4) Payment Service (e.g. Venmo, Cash App, Zelle, PayPal, KakaoPay, WeChat Pay), (5) Other}, 92 percent considered the use of a payment service provider, 4 percent chose credit card, and 4 percent "other".

Two final open-ended questions asked participants to share any scenarios in which they imagined a robot making a payment to a human, and any scenarios in which they imagined a human paying a robot. The survey also allowed participants to leave comments regarding the survey - considerations or reservations over the interaction with the robot.

Insights from the answers to the open-ended questions can be summarized as follows: 
\begin{itemize}
    \item The concept of money was mostly construed as fiat cash which is incompatible with the perception of robots
    \item The robot was mainly considered under a direct (master/slave) ownership model, where the owner of the robot meets the robot's financial obligation.
    \item Making payments to service robots and overall the model of Robot-as-a-Service (RaaS) was identified
    \item Few participants were able to propose scenarios in which a robot paid a human. The prevalent scenario was of a robot using a human utility such as electricity, or during constrained situations that render the robot immobile. 
\end{itemize}

Follow up interviews with selected participants identified that in that scenario, participants envisioned the interaction taking place with a humanoid robot. We consider this correlation an interesting social bias brought upon by the beliefs of our social structure and ignorance over the technology we present in this paper. This was noted after presenting the participants with a model of differential drive mobile-base robot. On the contrary - when presenting the same open-ended scenario to participants with knowledge in cryptocurrencies and blockchain, the interplay between blockchain technology and robotics perceived as fitting. Additionally, the latter group was able to propose more scenarios regarding payments and agreements between robots and humans.

Although this was a preliminary descriptive study, it serves as a starting point to further explore the human-robot interaction models afforded by blockchain technology and perceived biases over such interactions.

\section{Discussion}\label{discussion}

\subsubsection{Enforceability of Smart Contracts}
Legal enforceability of smart contracts and considerations over transacting with robots must be further explored by the community.

\subsubsection{Usability}
Blockchain technology has yet to achieve a mainstream user experience. Users of the technology must be able to securely store and manage private keys. Solutions such as storing private keys in hardware wallets are an option, or using software wallets and services such as \cite{2019_metamask} and \cite{2019_portis}. Users must become familiar with concepts like "gas" \cite{2014_buterin_ethereum} in the Ethereum network, transaction fees and other concepts. 


\subsubsection{Peer-to-peer Transactions}
If we trust that all parties will behave accordingly, not all transfer of value requires a smart contract. Consider the second scenario we presented in the introduction. If the robot employs a sophisticated vision recognition and event detection algorithm it could notice when a human subject has completed the task, and in turn reward the human subject. The robot could ask for the human's wallet address and send cryptocurrency directly.


\subsubsection{Future Applications}
Our current work explores the application of blockchain technology as part of a secure infrastructure for fully immersive telepresence robotics - robotic avatars that replicate an operator's motions and relay high-fidelity sensory feedback. Biometric information obtained by a telepresence control interface such as the one presented in \cite{2017_kim_engage_disengage} can be attached to a robot's decentralized identifier (DID) and linked to the DID of a human operator. Smart contracts can serve as access control bridges that require an operator's digital signature. Digital signatures can be also used to access control devices such as a telepresence control garment in \cite{2019_cardenas_telesuit_an_immersive_user_centric_telepresence_control_suit}. Additionally, we can consider the use of cryptocurrencies by robotic avatars to transact in a remote environment. Also work such as \cite{2019_unique_cipher_acoustic_languages}, which fingerprints interactions with robots, can leverage blockchain technology.


Lastly, we can consider that a RaaS model can leverage utility tokens to allow access to the service, and that real-time payments to robots could be made for the delivery of real-time services such as data collection.

\section{Conclusion}
This paper is complementary of a workshop presented at the 2019 RO-MAN conference. The software, including smart contracts and our open-source libraries used to connect robots to smart contracts can be found in \cite{2019_cardenas_ro_man_workshop}. In conclusion, the application of blockchain(s), smart contracts and cryptocurrencies in the field of robotics presents game changing opportunities. This encompasses the idea that a physical or virtual robotic agent can engage in, unmediated, peer-to-peer financial transactions with a human or other robots, as well as the idea of allowing robots to enter into agreements with humans and robots alike. When we envision the coexistence of robots and humans - blockchains, cryptocurrencies and smart contracts are key technologies that we must consider. Hence, this entails the combined study of human-robot interaction, behavioral economics, behavioral game theory and cryptoeconomics - from now on referenced as a whole as: Robonomics.

\bibliographystyle{plain}
\bibliography{references.bib}

\end{document}